\documentstyle[epsfig, twocolumn]{venice97}

\begin{document}

\setlength{\parindent}{0pt}
\setlength{\parskip}{ 10pt plus 1pt minus 1pt}
\setlength{\hoffset}{-1.5truecm}
\setlength{\textwidth}{ 17.1truecm }
\setlength{\columnsep}{1truecm }
\setlength{\columnseprule}{0pt}
\setlength{\headheight}{12pt}
\setlength{\headsep}{20pt}

\pagestyle{veniceheadings}

\title{\bf NEW RESULTS FOR THREE NEARBY OB ASSOCIATIONS}

\author{{\bf R.~Hoogerwerf$^1$, J.H.J.~de Bruijne$^1$, A.G.A.~Brown$^1$,
             J.~Lub$^1$, A.~Blaauw$^{1,2}$, P.T.~de Zeeuw$^1$}\vspace{2mm} \\
$^1$Sterrewacht Leiden, P.O.~Box 9513, 2300 RA Leiden, The Netherlands \\
$^2$Kapteyn Astronomical Institute, P.O.~Box 800, 9700 AV Groningen, The Netherlands}

\maketitle

\begin{abstract}

We have developed a new procedure to identify moving groups, and have
applied it to carry out a census of the nearby OB associations based
on Hipparcos parallaxes and proper motions (see de Bruijne et al.\ and
de Zeeuw et al.\ elsewhere in this volume). Here we present three
illustrative cases.  For $\alpha$ Persei our method allows us to
refine the bright end of the membership list, while for Collinder 121
the Hipparcos data change the whole appearance of the association.
Finally, we report the discovery of a new association in the field of
Cepheus~OB2.
\vspace {5pt} \\


Key~words: OB associations; moving groups.

\end{abstract}

\section{INTRODUCTION}

OB associations are dynamically unbound groups of young massive stars,
located near star forming regions. They form a perfect laboratory to
study the processes involved in high mass star formation in giant
molecular clouds (Blaauw 1964, 1991). Detailed knowledge on the
stellar content, i.e., the membership, provides information on the
initial mass function, the local star formation rate and efficiency,
differential age effects between subgroups within associations, the
binary population, and the interaction between stars and the
interstellar medium.

Membership of associations based on pre-Hipparcos proper motions is
known only for stars with spectral types earlier than B5 (e.g., Blaauw
1946; Bertiau 1958). Photometric studies extend membership to later
spectral types (e.g., Warren \& Hesser 1978; de Geus et al.\ 1989;
Brown et al.\ 1994). However, relying on the photometric distance as
the only membership criterion is dangerous: field stars are easily
selected as members due to the large physical size and loose structure
of an association.

To obtain reliable membership lists we have developed a new selection
procedure (de Bruijne et al.\ 1997) which uses Hipparcos data to
detect moving groups and determine membership. This work is part of a
larger investigation of the nearby OB associations and related star
forming regions (cf.\ de Zeeuw et al.\ 1994). Here we present three
examples of the results of the new member selection. An overview of
the project, and results for other associations, can be found in de
Zeeuw et al.\ (1997).

\section{$\alpha$ PERSEI}

Immediately after publication of the Preliminary General Catalog in
1910, Eddington, Boss, and Kapteyn independently reported the presence
of a moving group of B stars in the Perseus region. These stars have
large proper motions and stand clear from the field star population,
allowing for reliable membership selection. Rasmuson (1921) extended
the membership list with 45 stars of all spectral types. Heckmann et
al.~(1956, 1958) re-investigated the positions and proper motions in
the region of $\alpha$ Persei and found 163 members brighter than $V =
12.2$ down to spectral type G5. Much fainter members, $V < 19$, were
identified by Stauffer et~al.~(1985, 1989) and Prosser~(1992). These
studies resulted in about 300 known members. The $\alpha$ Persei
moving group is sometimes referred to as Per OB3, and may be related
to the nearby old Cas--Tau association (Blaauw 1956), which has
similar kinematics.

Our Hipparcos sample for $\alpha$ Persei consists of 425 stars of
spectral type earlier than G5 in the field $140^\circ\! < \ell <
155^\circ\!$ and $-11^\circ\! < b < -3^\circ\!$. Figure~1 shows the
vector point diagram, for the different spectral types. The moving
group is recognized easily in the B and A stars due to the small
velocity dispersion in the population of early-type field stars. The
separation between group and field becomes less clear for the F and G
stars because of the increased velocity dispersion in the field
population.

\begin{figure*}[!t]
  \begin{center} 
  \leavevmode
  \centerline{\epsfig{file=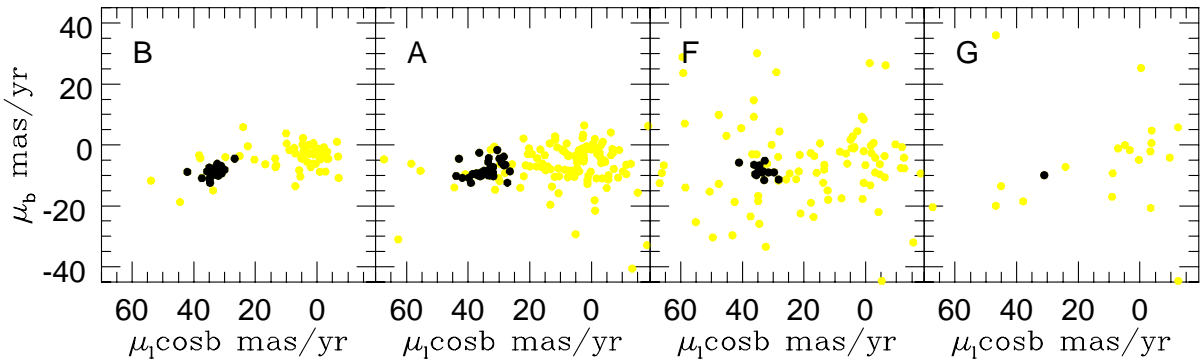,width=16.00truecm}}
  \end{center}
  \caption{\em Vector point diagrams for the B, A, F, and G stars in the
  $\alpha$ Persei field; black dots indicate members and gray dots are
  field stars. The separation between cluster and field is evident for 
  the B and A stars, becomes less clear for F stars, and has 
  disappeared for G stars.  
  }
\end{figure*}

We first apply our selection procedure to the B and A stars. Of the
254 stars 70 are found to be members. We then use the space motion of
the group defined by these B and A stars to search for comoving F and
G stars. Of the 171 F and G stars, 17 are found to be comoving with
the B and A members. In total we find 87 members; 32 B, 38 A, 15 F, 1
G, and 1 K type. The positions and proper motions are shown in
Figure~2. The apparent lack of late type members compared to e.g.,
Stauffer et al.\ (1985, 1987) and Prosser (1992) is caused by the
completeness limits of the Hipparcos Catalog. Of the 163 bright
members listed by Heckmann et al.~(1956) only 53 are contained in the
Hipparcos Catalog. Of these, 15 are rejected by our selection
procedure.

\begin{figure}[!b]
  \begin{center} 
  \leavevmode
  \centerline{\epsfig{file=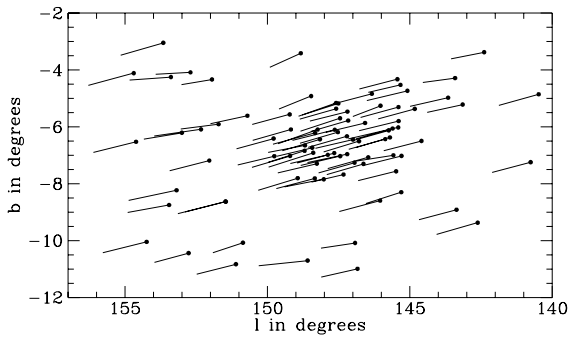,width=7.5cm}}
  \end{center} 
  \caption{\em Positions (dots) and proper motions (lines) of the
  $\alpha$ Persei members.}
\end{figure}

\begin{figure}[!b]
  \begin{center} 
  \leavevmode
  \centerline{\epsfig{file=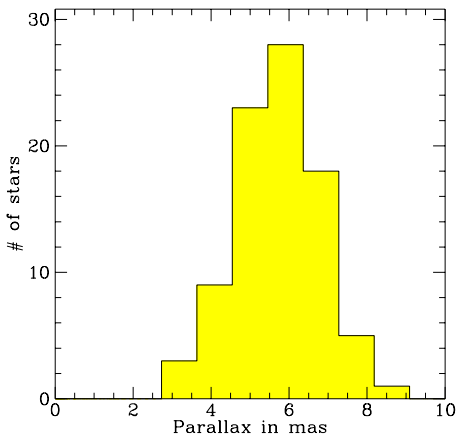,width=6.0cm}}
  \end{center} 
  \caption{\em Parallax distribution of the $\alpha$ Persei
               members. The mean parallax is $5.7 \pm 0.1$ mas.}
\end{figure}

\begin{figure*}[!t]
  \begin{center} 
  \leavevmode
  \centerline{\epsfig{file=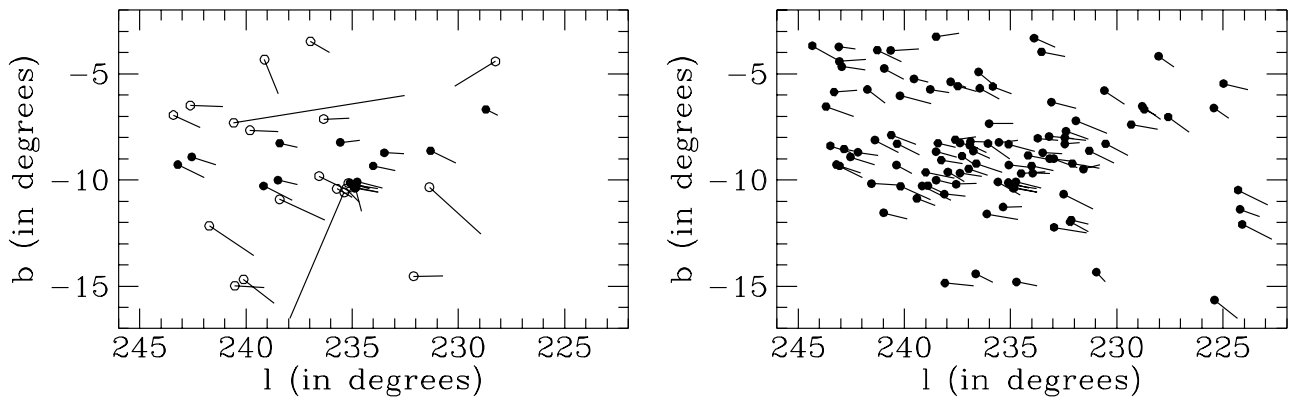,width=\textwidth}}
  \end{center} 
  \caption{\em Left: Hipparcos positions and proper motions of the classical
  `members' of Collinder 121. Filled dots are stars confirmed as member,
  while open dots are rejected by our selection procedure. Right: same
  plot for the stars we select as members of Collinder 121. Note the
  dramatic changes from pre-Hipparcos to post-Hipparcos.}
\end{figure*}

Following the procedure described in de Zeeuw et~al.~(1997) we obtain
a mean distance of $176 \pm 5$~pc for the $\alpha$ Persei cluster,
where the error corresponds to the error in the mean parallax. This
agrees with previous distance estimates which range from 160 to 180 pc
(e.g., Roman \& Morgan 1950; Mitchell 1960). The parallax distribution
(Figure~3) shows that the cluster has a depth less than $\sim 60$ pc,
but with an average error of 1 mas in parallax we are unable to
resolve the depth of the cluster.

\section{COLLINDER 121}

Collinder 121 was first recognized as a physical group of stars by
Collinder (1931). He found an open cluster of 20 stars at a distance
of 590 pc within an area of $1^\Box$. Of these, 8 are contained in the
HD catalog. Feinstein (1967) found 11 stars belonging to the cluster
in the central region defined by Collinder, and he extended the
membership list with 16 bright and 12 faint stars in a $10^\circ$
circle around the cluster center. Feinstein put the cluster at 630
pc. Eggen (1981) suggested a division into two groups for the stars in
this region: an open cluster-like group of 13 stars at 1.17 kpc and
another group of 18 stars at 740 pc resembling an OB association.
Figure~4 shows the stars listed by Feinstein as classical members of
Collinder 121.  Evidence for a moving group can be seen in the
Hipparcos proper motions, although some of the stars are clearly
non-members based on the magnitude and direction of the proper motion.

\begin{figure}[!b]
  \begin{center} 
  \leavevmode
  \centerline{\epsfig{file=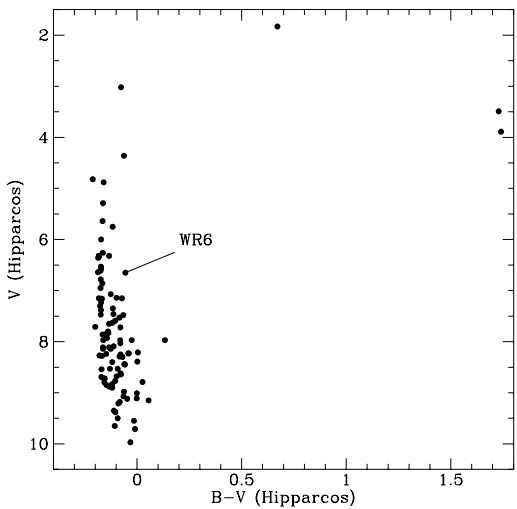,width=6.8cm}}
  \end{center} 
  \caption{\em Colour magnitude diagram, uncorrected for reddening,
  for the Collinder 121 members. The earliest spectral type is B2. The
  unusual position of WR6 is caused by systematic effects in the $Hp$
  to $V$ and $B_T - V_T$ to $B - V$ transformation. Feinstein (1967)
  obtained $V = 6.91$, $B-V = -0.28$.}
\end{figure}

Our Hipparcos sample for Collinder 121 consists of all O and B stars
in the field $222^\circ\! < \ell < 246^\circ\!$ and $-16^\circ\! < b <
-3^\circ\!$. After applying our selection procedure we are left with
105 of the 449 stars. The earliest spectral type among the members is
B2 (Figure~5). Most of the members are evolved early-type stars
suggesting an old association. Collinder 121 has dramatically changed
its appearance compared to the previous membership lists (cf.\
Figure~4). While 5 of the 8 HD stars listed as members by Collinder
are contained in the Hipparcos Catalog, only 2 are selected as
member. From the 39 stars listed by Feinstein, 35 are in the Hipparcos
Catalog and 10 are classified as members by our selection
procedure. For Eggen's distant group in the Collinder 121 region we
select 4 from the 7 stars contained in the Hipparcos Catalog, and
another 3 from the 6 stars in the closer group. There is no clear
difference in the parallaxes for these two groups, suggesting that the
division made by Eggen is unphysical.

We derive a distance of $546 \pm 30$ pc for Collinder 121. This is
closer than all previous distance determinations.  The well-studied
Wolf-Rayet star WR6 (EZ CMa) is a member. Its parallax of $1.74 \pm
0.76$ mas places the star in the middle of the cluster. This is in
disagreement with the lower limit on the distance of 1.8 kpc derived
by Howarth \& Schmutz (1995), and indicates that the absolute
luminosity of this WR star is an order of magnitude lower than thought
previously.

\section{A NEW GROUP}

We have discovered a previously unknown group of comoving early-type
stars in the Cepheus~OB2 field. The Hipparcos sample consists of 388
stars (all B stars, plus all A stars fainter than $V = 8$) in the
field $96^\circ\! < \ell < 108^\circ\!$ and $-1^\circ\! < b <
12^\circ\!$. The new group contains 15 stars with spectral types later
than B6, and is located at the edge of the Cepheus~OB2 field. Their
positions and proper motions clearly reveal a moving group (cf.\
Figure~6). Figure~7 shows the Hipparcos colour-magnitude diagram. The
main sequence is evident, which suggests the stars are at a similar
distance. The lack of early spectral types suggests an old
association, similar to Collinder 121. However, we may well have
missed the most massive members of the group due to the field limits
(Figure~6). A more detailed description of this group will be possible
when the whole Hipparcos Catalog becomes available, and we can search
the adjacent area. We have initiated a radial velocity study to refine
the membership list further.

The IRAS 60 micron map for the Cepheus region shows a varying
intensity across the field of this new group.  The scatter around the
main sequence may thus be caused by variable extinction.  Str\"omgren
photometry is available for one of the two stars above the main
sequence, and indicates a reddening larger than 0.5 magnitudes in~$V$.
It will be interesting to obtain homogeneous multi-colour photometry
for all members of this group.

We derive a distance of $240 \pm 10$ pc for the new group. This places
it inside the Gould Belt, the well-known ring-like distribution of
early-type stars which is tilted by $\sim 20^\circ\!$ with respect to
the Galactic plane. Some of the nearby OB associations, including
Sco~OB2, Ori~OB1 and Per~OB2, are part of this structure (e.g.,
Lindblad et al.\ 1997; Torra et al.\ 1997, and references
therein). The $\alpha$ Persei cluster and the Cas--Tau association are
located in the middle of the Gould Belt, and may be related to its
origin.

\section{CONCLUDING REMARKS}

\begin{figure}[!t]
  \begin{center} 
  \leavevmode
  \centerline{\epsfig{file=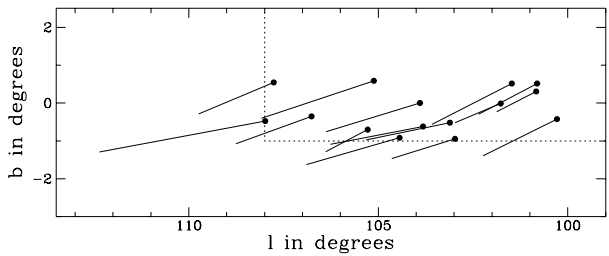,width=8.0cm}}
  \end{center} 
  \caption{\em Positions and proper motions for members of a new group
  of comoving stars in the Cepheus~OB2 field. The dotted lines
  are the field limits of our present sample.}
\end{figure}

We have presented results on the member selection and distance
analysis of two known OB associations, $\alpha$ Persei and
Collinder~121, and one newly discovered group in the field of
Cepheus~OB2. They are part of a larger census of the nearby OB
associations discussed by de Zeeuw et al.\ elsewhere in this
volume. The three examples given here illustrate the major step
forward in our understanding of OB associations provided by
Hipparcos. We refined the membership list for $\alpha$ Persei, based
on 87 bright members identified in our Hipparcos sample.  The accuracy
of the membership list was already good due to the large tangential
velocity of $\alpha$ Persei, 30 km/sec with respect to the Sun.  The
new distance of $176 \pm 5$ pc agrees well with previous estimates.
The appearance of Collinder 121 changed from a compact open cluster
($1^\Box$, 20 members at 590 pc) to an old OB association at $546 \pm
30$ pc with 105 members distributed over $\sim45^\Box$.  Finally, we
discovered a new group of 15 comoving early-type stars in the
Cepheus~OB2 region at $240 \pm 10$ pc. Further investigation is needed
because the group is located at the edge of the field and we expect to
find more members when the whole Hipparcos Catalog becomes available.

\section*{ACKNOWLEDGMENTS}

It is a pleasure to thank Michael Perryman and Rudolf Le Poole for
many stimulating discussions.

\begin{figure}[!t]
  \begin{center} \vskip -3.0truemm \leavevmode
  \centerline{\epsfig{file=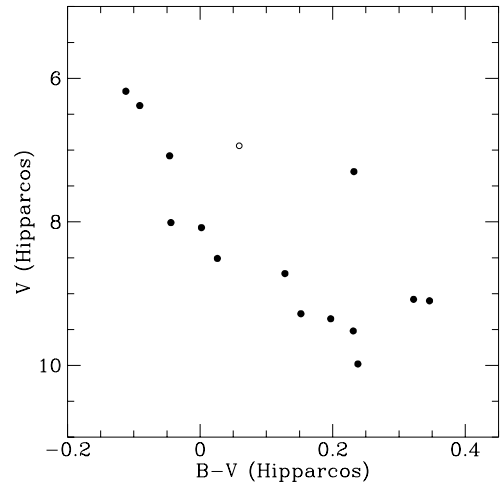,width=6.5cm}}
  \end{center} \caption{\em Colour magnitude diagram, uncorrected for
  reddening, for members of the new group. The earliest spectral type
  is B6. The scatter of the stars around the main sequence is probably
  due to varying extinction across the field. The open dot is a star
  with an extinction of more than 0.5 magnitude, putting it close to
  the main sequence.}
\end{figure}

\end{document}